\documentclass[12pt,a4paper]{scrartcl}

\usepackage{graphicx}
\graphicspath{{figures/}}
\usepackage{amsmath}
\usepackage{amssymb}
\usepackage{xcolor}
\usepackage[T1]{fontenc}
\usepackage{textcomp} 
\usepackage[auth-sc-lg, affil-it]{authblk} 

\usepackage[implicit=true,colorlinks=true,linkcolor=blue,citecolor=blue,urlcolor=blue]{hyperref}
\usepackage[numbers,sort&compress]{natbib}

\newcommand{\units}[1]{\ensuremath{\,\mathrm{#1}}}
\newcommand{\SE}{Schr\"odinger equation}

\newcommand{\dd}{\mathrm{d}}
\newcommand{\ii}{\mathrm{i}}
\newcommand{\ee}{\mathrm{e}}

\begin{document}

\title{From Newton's equations to the Schr\"odinger equation: domain walls as the quantization condition}

\author[1]{Dennis M. Heim\thanks{physics@d-heim.de}}
\affil[1]{67663 Kaiserslautern, Germany}
\maketitle

\begin{abstract}
    Can particles at definite positions in real space, obeying Newton's equations under an ordinary interaction force, have a coarse-grained description that is the \SE{}? We present a medium of \emph{companion particles} on a line under a force of exactly that kind: each companion pays for the mismatch of the signed inverse gaps to its two neighbors. A binary polarity carried by each companion makes stationary states with nodes dynamically stable and supplies the quantization condition. Finite chains relax into the stationary densities of the box and the harmonic oscillator and develop double-slit fringes. In the smooth limit the \SE{} emerges, with $\psi$ carrying two real fields rather than being fundamental.
\end{abstract}

\section{Introduction}

Attempts to restore definite particle positions beneath quantum mechanics are as old as the theory itself. Madelung rewrote the Schr\"odinger equation as the hydrodynamics of a density and a velocity field~\cite{madelung:1926}; de~Broglie and Bohm supplemented it with particles guided by the wave function~\cite{debroglie:1927, bohm:1952:1, bohm:1952:2, bohm:1975}; Nelson derived it from a stochastic process in configuration space~\cite{nelson:1966}. Each recovers standard predictions, and each remains under active discussion~\cite{wallstrom:1994, nauenberg:2014, passon:2018, reddiger:2023, valentini:2025}. What they share is structural: none is Newtonian in the plain sense. Bohm's particles do not obey $F = ma$ but a first-order guidance law driven by a field on configuration space; Nelson's obey a diffusion whose random force has no mechanical origin; Madelung's fields are a rewriting of $\psi$, not an account of what underlies it.

This suggests a sharper question: Can particles that are localized at definite positions in real space, obeying Newton's equations under an ordinary interaction force, have a coarse-grained description that is the \SE{}?

This paper presents a model of that kind. Alongside the particle an experimenter tracks, it places a medium of \emph{companion particles} on the line: particles at definite positions, interacting only with their nearest neighbors as a particle of mass $m$ would. Each companion carries, besides its position, a binary \emph{polarity} $s_k \in \{+1,-1\}$. The interaction is built from the signed inverse gaps to the two neighbors. A single coupling constant fixes its strength.

Smoothing the chain into a density and a velocity field, and asking only that the companions be conserved, yields the \SE{}, with $\psi$ appearing as a compact way to carry two real fields rather than as a fundamental object. Integrated directly under Newton's equations, chains of a few dozen to a few hundred companions relax into the stationary densities of the particle in a box and of the harmonic oscillator and develop double-slit fringes.

The interaction this construction arrives at has been written down before. In the \emph{many interacting worlds} (MIW) model of Hall, Deckert, and Wiseman~\cite{hall:2014}, a large but finite number of classical ``worlds'' repel each other through a potential built from the inverse spacings between neighboring world configurations, and the ensemble reproduces Ehrenfest's theorem, wavepacket spreading, tunneling, zero-point energy, oscillator ground states, and double-slit interference. The same program, without a discrete ensemble, appears in the labeled continuum of trajectories of Poirier and of Schiff and Poirier~\cite{poirier:2010, schiff:2012}, and in Sebens's collection of Newtonian worlds, whose force law is written in the continuum~\cite{sebens:2015}.

Wherever the polarity is uniform, the interaction of Section~\ref{sec:interaction} is equal to the MIW interworld potential. However, the two constructions reach it by different routes. MIW begins with the quantum average energy and discretizes it over an ensemble of worlds, so the correspondence with quantum mechanics is built in from the start. The present construction begins with an ordinary force between particles on a line.

Two results are new here.
\begin{itemize}
    \item \textbf{A binary polarity supplies the quantization condition.} The polarities make the stationary states with nodes dynamically stable. A finite chain of binary labels carries an integer number of domain walls, and that count labels the state (Section~\ref{sec:examples}).
    \item \textbf{Energy transitions have a mechanism.} A transition does not start in the bulk: a wall enters at a boundary and migrates inward one flip at a time, so the medium prefers resonant over impulsive excitation (Section~\ref{sec:walls}).
\end{itemize}

\section{The Microscopic Interaction}
\label{sec:interaction}

\subsection{Setting the Scene}

Picture a line populated by many particles, each with a definite position $x_k(t) \in \mathbb{R}$. One of them is \emph{visible}: the one an experimenter tracks. The rest are \emph{companion particles}: real in occupying definite positions and obeying the same equation of motion, but carrying no mass of their own. Each acts on its neighbors as a particle of mass $m$ would, without being one. Only the tracked particle is matter, the medium is the structure it moves in. What interaction among the companions could give rise to the behavior we associate with quantum mechanics?

\subsection{Three Particles}
\label{sec:three-particles}

Start with three particles at positions $x_1 < x_2 < x_3$. The middle one has the gap
$x_2-x_1$ on its left and $x_3-x_2$ on its right, and we ask for an interaction potential
that vanishes when the two gaps agree, grows as they differ, and diverges as either one
closes. A simple choice is
\begin{equation}
    V \;\propto\; \left(\frac{1}{x_3-x_2} - \frac{1}{x_2-x_1}\right)^{2}.
    \label{eq:v-three-unpolarized}
\end{equation}
Each inverse gap is the local particle density on that side, so the subtraction measures
the density imbalance across the middle particle: the medium favors uniform density
without committing to a preferred spacing. A gap that closes while its partner stays
open costs an unbounded energy.

To make the medium able to form domains separated by domain walls, we give each companion a \emph{binary polarity} $s_k \in \{+1,-1\}$. A domain is a stretch of companions with the same polarity. A link between opposite polarities is a domain wall. The three-particle energy then becomes
\begin{equation}
    V \;\propto\; \left(\frac{s_2s_3}{x_3-x_2} - \frac{s_1s_2}{x_2-x_1}\right)^{2}.
    \label{eq:v-three-polarized}
\end{equation}

\subsection{The Chain}

Every particle inside the chain makes the same comparison between the signed inverse gaps on its left and right. Adding all these squared differences gives the total interaction energy
\begin{equation}
    V_\mathrm{int} \;=\; \kappa \sum_k\!\left(\frac{s_k\, s_{k+1}}{x_{k+1}-x_k}
    \;-\; \frac{s_{k-1}\, s_k}{x_k-x_{k-1}}\right)^{\!2}.
\end{equation}
The constant $\kappa$ sets the strength of the interaction. The terms inside the square have units of inverse length, so $\kappa$ must have units of energy $\times$ length-squared. The combination $\hbar^2/m$ has exactly those units, so we write $\kappa = \hbar^2/(8m)$, with $\hbar$ a constant of the medium whose size a single experiment fixes. This leads to the interaction potential
\begin{equation}
    \boxed{\;
        V_\mathrm{int}
        \;=\; \frac{\hbar^2}{8m}\sum_k\!\left(\frac{s_k\, s_{k+1}}{x_{k+1}-x_k}
        \;-\; \frac{s_{k-1}\, s_k}{x_k-x_{k-1}}\right)^{\!2}.
        \;}
    \label{eq:vint}
\end{equation}

With uniform polarity all $s_k s_{k+1} = 1$, and the interaction \eqref{eq:vint} is precisely the interworld potential of the MIW model~\cite{hall:2014}. There it is reached by discretizing the quantum average energy over an ensemble of world configurations. The polarity is the ingredient it does not have: the stability of the eigenstates (Section~\ref{sec:examples}) and the mechanism of the transitions between them (Section~\ref{sec:walls}) rest on it.

\section{Examples}
\label{sec:examples}

The interaction \eqref{eq:vint} defines an ordinary classical many-body problem: $N$ companion particles on a line, moving under Newton's equations with forces obtained from the negative gradient of the discrete interaction potential, plus any external potential $V(x)$. The question is whether \emph{finite} chains, here $N = 60$ to $300$, can reproduce the predictions of the \SE{}. This section presents three standard tests: the particle in a box, the harmonic oscillator, and the double slit. All simulations use second-order (Verlet-type) integration with an adaptive time step, nanometer/picosecond units, and the electron mass.

Stationary states are obtained by \emph{damped relaxation}. Starting at rest from a uniformly spaced chain, the companion particles move under
\begin{equation}
    m\,\ddot{x}_k \;=\; -\frac{\partial V_\mathrm{tot}}{\partial x_k}
    \;-\; m\,\gamma\,\dot{x}_k,
    \qquad
    V_\mathrm{tot} \;=\; V_\mathrm{int} + \sum_j V(x_j),
    \label{eq:relax}
\end{equation}
with the constant friction $\gamma = 0.4\units{ps^{-1}}$. The friction removes the kinetic energy of the transient and the chain settles into a minimum of $V_\mathrm{tot}$. The relaxed densities $P_k = 1/(x_{k+1}-x_k)$, read at the link midpoints and normalized to their maximum, are then compared with the density $|\psi|^2$ of the corresponding stationary solution of the \SE{}. The double slit is instead evolved without damping and compared with the time-dependent solution.

Each example also needs a rule for what lies beyond the ends of the chain, because the sum in Eq.~\eqref{eq:vint} reaches one link past the last particle. In every case the chain is extended, so that the same sum runs unchanged over the extension. Two extensions suffice: reflection in a \emph{hard wall}, with the image polarities flipped, and \emph{vacuum}, in which the gaps to the absent neighbors are taken infinite. That is, the inverse gaps are zero. The box uses the first, the harmonic oscillator and the double slit use the second.

\subsection{Particle in a Box}
\label{sec:example-box}

The first test is the standard particle in a box: hard walls at $x = \pm b$ with $b = 50\units{nm}$ and $V = 0$ in between. Each wall is implemented by reflecting the chain in it and giving the image particles the opposite polarity, so that a wall acts as a domain wall of the medium, pinned at $\pm b$.

\begin{figure}[t!]
    \centering
    \includegraphics[width=\linewidth]{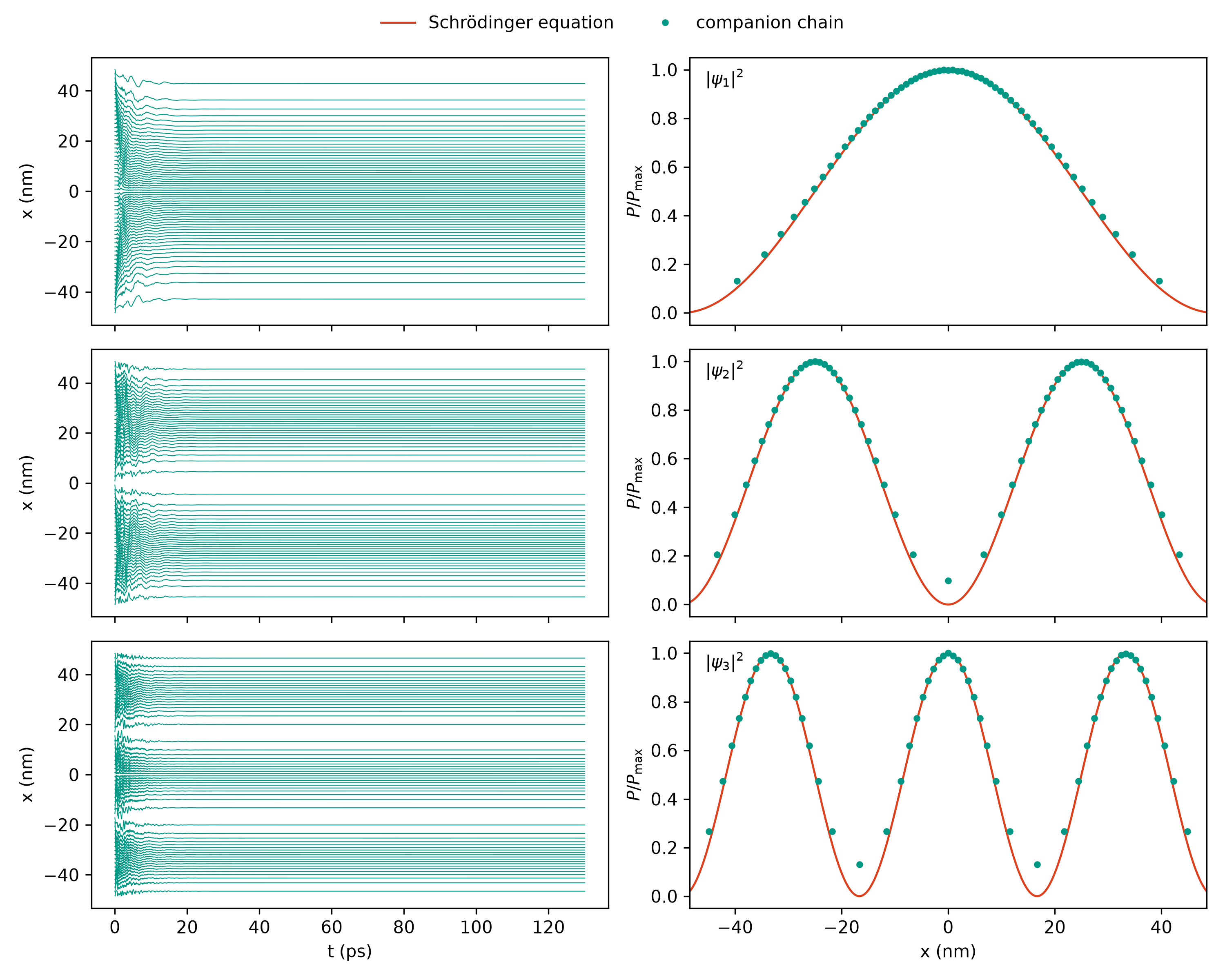}
    \caption{Particle in a box ($b = 50\units{nm}$): damped relaxation of chains of $N = 60$ companion particles to the three lowest states $n = 1, 2, 3$ (top to bottom). Left: particle trajectories $x_k(t)$ from a uniformly spaced start. Right: final normalized link densities $P/P_\mathrm{max}$ (dots) against the stationary densities of the \SE{} (solid line).}
    \label{fig:box}
\end{figure}

Figure~\ref{fig:box} shows damped relaxations to the three lowest states from uniformly spaced starts: all polarities equal (ground state), a single flip between the two central particles (one domain wall), and flips after particles 20 and 40 (two walls, the three domains hold 20 particles each). The relaxed densities agree well with the eigenstates of the \SE{}, apart from the node links, where a finite gap carries a small residual density.

Here the flips were placed by hand. Section~\ref{sec:walls} discusses how the interaction~\eqref{eq:vint} itself can explain their positions and what moving them costs.

\subsection{Harmonic Oscillator}
\label{sec:example-hosc}

The harmonic oscillator replaces the hard walls by a smooth confinement, $V = \tfrac{1}{2} m \omega^2 x^2$ with $\omega = 2\pi/60\units{ps^{-1}}$, for which the \SE{} predicts stationary densities $|\psi_n|^2$ of Gaussian scale $\sigma = \sqrt{\hbar/(2m\omega)} = 23.51\units{nm}$. Unlike in the box, the ends of the chain are open, with vacuum beyond them. The interaction pushes the end of the chain outward against the vacuum, the external force pulls it back, and their balance lies beyond the classical turning point.

\begin{figure}[t!]
    \centering
    \includegraphics[width=\linewidth]{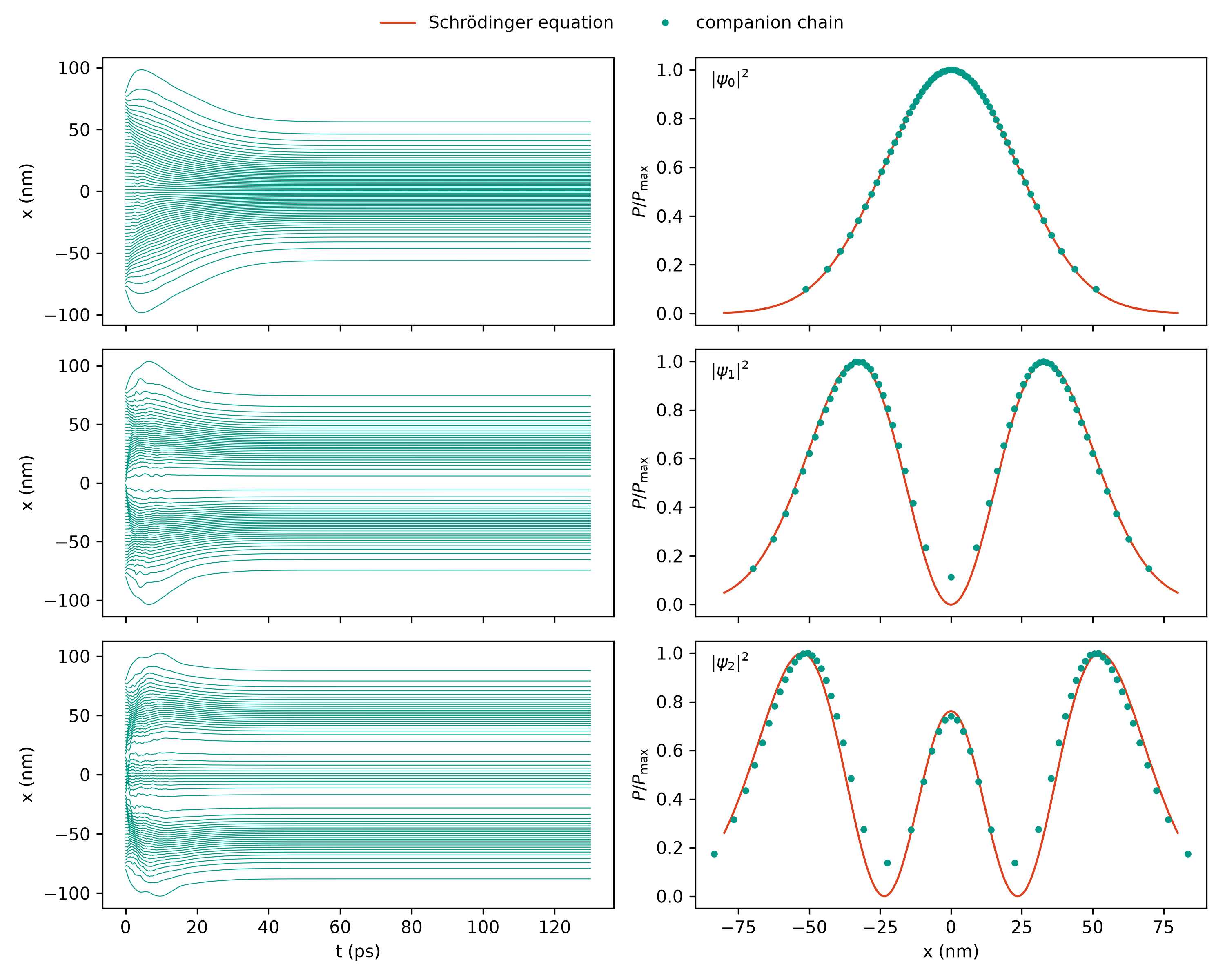}
    \caption{Harmonic oscillator ($\omega = 2\pi/60\units{ps^{-1}}$): damped relaxation of chains of $N = 60$ companion particles to the states $n = 0, 1, 2$ (top to bottom). Left: particle trajectories $x_k(t)$. Right: final normalized link densities $P/P_\mathrm{max}$ (dots) against the stationary densities of the \SE{} (solid line).}
    \label{fig:hosc}
\end{figure}

Figure~\ref{fig:hosc} shows the relaxed chains with $N = 60$ particles for $n = 0$ (all polarities equal), $n = 1$ (one central flip) and $n = 2$ (flips positioned so that the domains hold $24/12/24$ particles). As in the box, the relaxed densities agree well with the stationary densities of the \SE{}, with a small residual density at the node links. These divisions are the shares of that stationary density between its nodes. They were set by hand here. Section~\ref{sec:walls} shows for the box that the interaction itself makes the walls sit at such shares.

\subsection{Double Slit}
\label{sec:example-dslit}

Electron double-slit diffraction~\cite{bach:2013} probes the dynamics rather than the stationary states. The initial density consists of two Gaussians of width $10\units{nm}$ centered at $\pm 50\units{nm}$. A total of $N = 300$ particles are placed at equal quantiles of this profile and released at rest, with $V = 0$ and vacuum beyond the chain ends. No polarity variation enters. The chain evolves undamped for $20\units{ps}$.

\begin{figure}[t!]
    \centering
    \includegraphics[width=\linewidth]{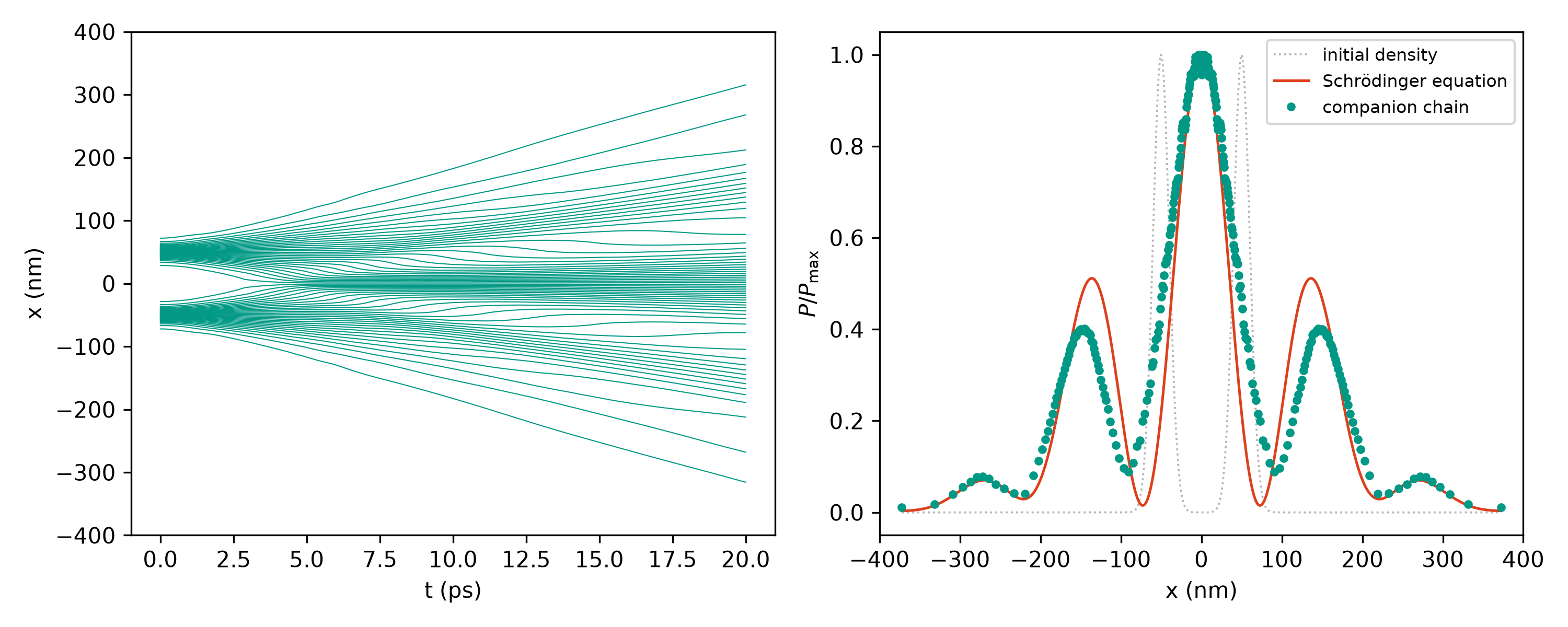}
    \caption{Double slit: undamped time evolution of $N = 300$ companion particles released at rest from two Gaussians of width $10\units{nm}$ at $\pm 50\units{nm}$. Left: particle trajectories $x_k(t)$, every fifth companion shown. The packets spread, overlap, and organize into fringes. Right: normalized link density at $t = 20\units{ps}$ (dots) against $|\psi(x,t)|^2$ of the \SE{} for the same initial condition (solid line) and the initial density (dotted).}
    \label{fig:dslit}
\end{figure}

Figure~\ref{fig:dslit} compares the evolved chain with $|\psi(x,t)|^2$ from the \SE{} for the same initial condition. The fringes emerge at the right positions and in the right number: the two clouds spread, overlap, and organize themselves under the interaction~\eqref{eq:vint}, with nothing wave-like put in. However, the first side maxima come out too low, which should be a topic of future investigations.

\section{Domain Walls and Transitions}
\label{sec:walls}

The states of Section~\ref{sec:examples} were prepared by hand: the number of domain walls and the links carrying them were set at the start. In the following we show how the interaction~\eqref{eq:vint} can explain the formation of these positions and how much energy the transition from one stationary state to another costs.

\subsection{Where a New Wall Can Form}

Each companion compares the link on its left with the link on its right. Along a smooth uniform-polarity chain the two nearly cancel: the differences in~\eqref{eq:vint} stay small, and so does the energy. Inserting a domain wall turns two of these differences into sums, and it is these sum terms that raise the energy, in proportion to the square of the inverse gaps there.

A wall is therefore cheapest where the medium is already thin, and the thinnest place is a boundary. This has a magnetic analogue: in many ferromagnets, reversal nucleates preferentially at surfaces, edges, inclusions, or defects, where the barrier is reduced~\cite{brown:1945, hubert:1998}. In the companion medium, the wall enters at the boundary and migrates inward, one flip at a time.

\subsection{What a Domain End Stores}
\label{sec:domain-ends}

A domain wall in a magnetic medium carries an energy of its own~\cite{landau:1935, hubert:1998}. The model postulates the same here: each domain end stores an energy fixed by the link that closes it and by the domain itself,
\begin{equation}
    \boxed{\;
        E_\mathrm{end} \;=\; \frac{\hbar^2}{2\pi m}
        \left(\frac{1}{x_{k+1}-x_k}\right)^{\!2}
        \left(1 + \frac{4}{N_\mathrm{d}^{2/3}}\right),
        \;}
    \label{eq:wall-energy}
\end{equation}
with $N_\mathrm{d}$ the number of companions in the domain it closes off. Each of the two domains meeting at a wall stores one such amount. At a hard wall the chain is closed by its mirror image, and the gap to that image plays the same role. A state of $n$ domains therefore carries $2n$ ends.

\subsection{From the First State to the Second}

Pulling a wall from the boundary to the center means first flipping the outermost inner particle's polarity. The mirror particle is flipped automatically by the boundary convention. Then the next particle inward is flipped too, and so on. Each flip costs energy as the chain adapts by widening gaps. The cost rises monotonically from the boundary, where the box wall already supplies the domain wall, to the center, where it reaches the second stationary state.

Because companions cannot cross, the wall's position fixes the share of the medium on each side. At a given share the chain settles to its cheapest arrangement: each lobe grows cheaper the more length it is given, and the lobes compete for the length available. A wall feels no force only where this cheapest energy is stationary in the shares. At this endpoint the two lobes have equal cheapest energies and the chain has reached the second solution of the \SE{}.

\begin{figure}[t!]
    \centering
    \includegraphics[width=\linewidth]{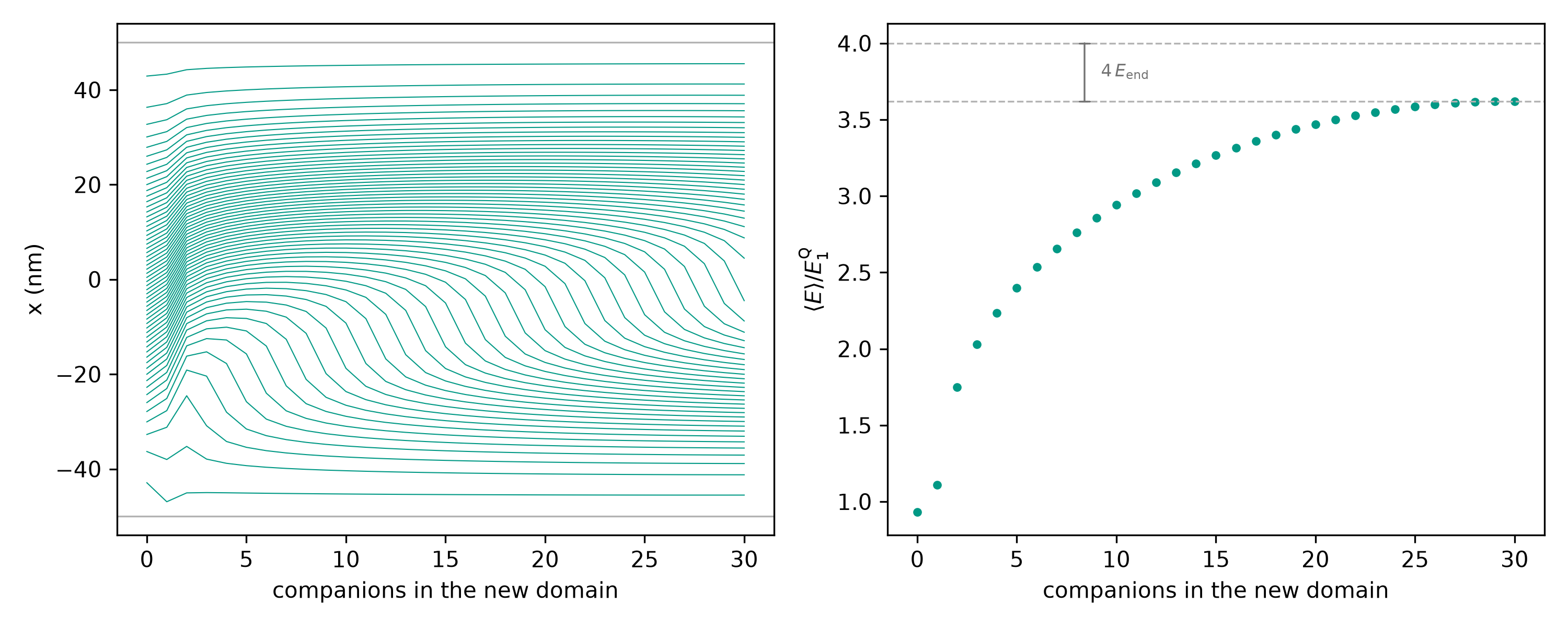}
    \caption{Ground state to first excited state ($b = 50\units{nm}$, $N = 60$). Left: relaxed companion positions as the wall is held at one link after another, moving inward from the boundary. Gray lines mark $\pm b$. Right: average interaction energy of the relaxed chain in units of $E_1^\mathrm{Q}$, the energy the \SE{} gives in the ground state. Dashed lines mark the chain's plateau and $E_2^\mathrm{Q}/E_1^\mathrm{Q} = 4$. Their gap is $4E_\mathrm{end}/N$: the energy~\eqref{eq:wall-energy} of the four domain ends of that state, two at the wall and one at each box wall, per companion. The climb starts below $1$ for the same reason: the ground chain's own two ends are not part of its interaction energy. The horizontal coordinate is the number of companions in the new domain.}
    \label{fig:transition}
\end{figure}

Figure~\ref{fig:transition} shows this quasi-static ground-to-second-state climb, computed by holding the wall at each link in turn and relaxing the chain at the resulting shares. The energy shown is the interaction energy per companion: each companion contributes one term of the interaction~\eqref{eq:vint}, and these terms are averaged over which companion is the tracked massive particle. The unit is $E_1^\mathrm{Q}$, the energy the \SE{} gives in the ground state. The climb is monotone and plateaus at $3.62$, below the $4$ of the \SE{} by the four ends the second state carries. It starts not at $1$ but at $0.93$, since the ground chain too falls short of $E_1^\mathrm{Q}$ by what its own two ends store. Moving the wall to the right or left would lower the total energy, but would also lead to a collapse into the ground state.

\subsection{From the Second State to the Third}
\label{sec:second-to-third}

\begin{figure}[t!]
    \centering
    \includegraphics[width=\linewidth]{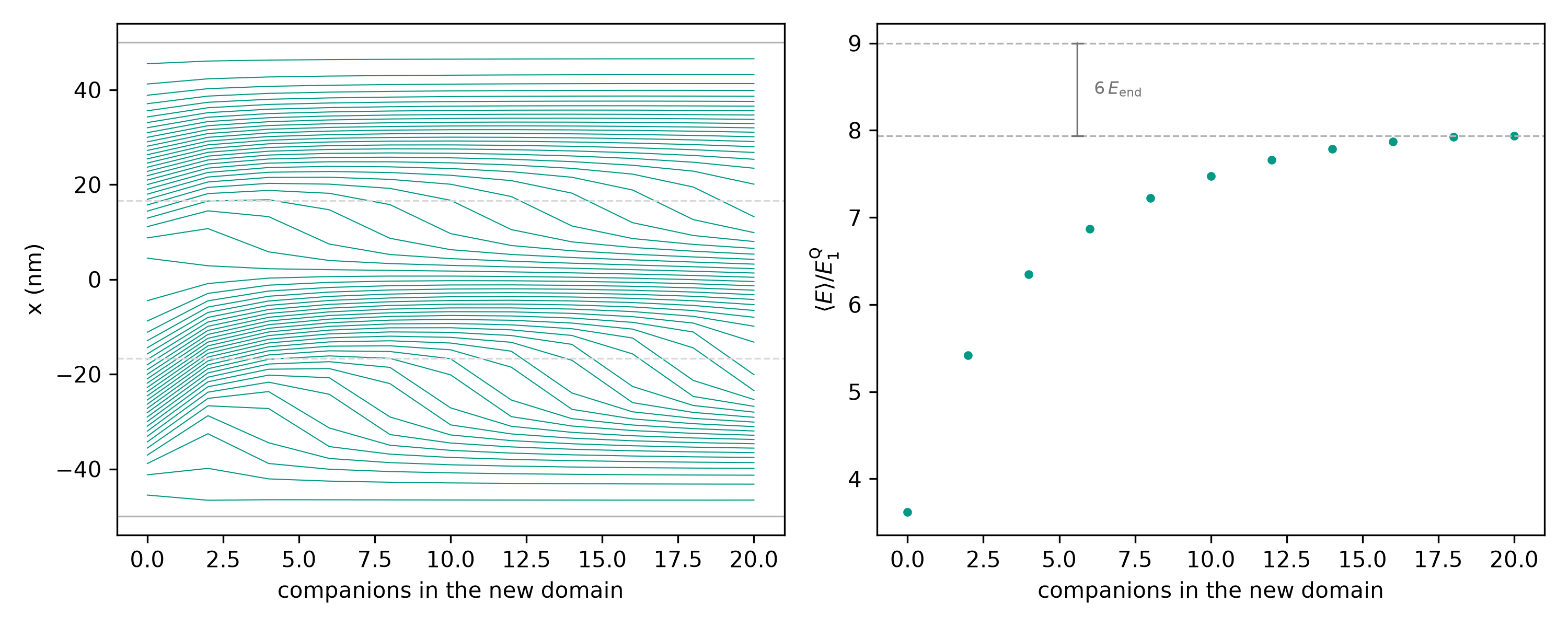}
    \caption{Second state to third state ($b = 50\units{nm}$, $N = 60$). Left: a new wall enters at $-b$ while the existing wall migrates, reaching $\pm b/3$ (dashed) together. Gray lines mark $\pm b$. Right: average interaction energy in units of $E_1^\mathrm{Q}$. Dashed lines mark the chain's plateau and $E_3^\mathrm{Q}/E_1^\mathrm{Q} = 9$. Their gap is $6E_\mathrm{end}/N$: the energy~\eqref{eq:wall-energy} of the six domain ends of that state, two at each of the two walls and one at each box wall, per companion. The horizontal coordinate is the number of companions in the new domain.}
    \label{fig:transition-23}
\end{figure}

The second state is divided into halves, and no new wall alone can divide it into thirds: since companions cannot cross, the existing wall must migrate as well, changing the shares along
\[
    (0,\tfrac12,\tfrac12) \longrightarrow (\tfrac13,\tfrac13,\tfrac13).
\]
Figure~\ref{fig:transition-23} shows this route: a second wall enters at $-b$ while the existing wall migrates, the two reaching $\pm b/3$ together. The climb is again monotone and carries the chain on from the $3.62$ of Figure~\ref{fig:transition} to $7.94$, below the $9$ of the \SE{} by the six ends of the third state. Higher transitions follow the same pattern, a new wall entering from a boundary while the others migrate.

\subsection{What a Transition Costs}

With the end energies counted, the relaxed box states carry the energies the \SE{} gives, and a transition costs the difference between two such totals.

The chain relaxes with the wall at any link, but feels no force only at the stationary divisions. The computed barrier is moreover a quasi-static minimum: with no excess energy available the transition must be driven slowly and coherently across the medium, whereas an impulse excites other motion and fails to open the gap. The resulting preference for resonant over impulsive excitation is a prediction of the model.

\section{From Companion Particles to the Schr\"odinger Equation}
\label{sec:schrodinger}

\subsection{Smoothing the Interaction}
\label{sec:smoothing}

Consider a stretch of the chain over which the polarity is uniform, so that the signed link quantities in the interaction energy~\eqref{eq:vint} reduce to the bare inverse gaps $P_k \equiv 1/(x_{k+1}-x_k)$. Where these gaps are short compared with the distance over which $P_k$ itself changes appreciably, the discrete distribution $\{x_k\}$ is well described by a smooth density $P(x,t)$, with $P_k \to P(x_k, t)$ and
\begin{equation}
    P_k - P_{k-1} \;\approx\; (x_k - x_{k-1})\, \partial_x P \;=\; \frac{\partial_x P}{P}.
    \label{eq:smooth-step}
\end{equation}
The replacement is controlled by the dimensionless ratio $|\partial_x P|/P^2$, the change of the density over one gap relative to the density itself, and is valid where this ratio is small.

Summing the squared differences~\eqref{eq:smooth-step} over the companions, one per length $1/P$, turns the discrete sum into the integral
\begin{equation}
    V_\mathrm{int} \;\longrightarrow\; \frac{\hbar^2}{8m}\int \dd x\, \frac{(\partial_x P)^2}{P}.
    \label{eq:vint-smooth}
\end{equation}
This is the smooth form of the companion-particle interaction energy. It depends only on the density profile $P(x,t)$ and vanishes precisely when $P$ is uniform: the continuum expression of the companions' aversion to density gradients.

The functional on the right of Eq.~\eqref{eq:vint-smooth} has appeared before. It is the von Weizs\"acker kinetic-energy functional, introduced in nuclear-mass theory as a density-gradient correction to the local Fermi-gas kinetic energy~\cite{weizsacker:1935}. Up to a constant factor, it is the Fisher information of the density $P$. It also enters several derivations of the \SE{}, but by distinct routes: Reginatto introduces it through a principle of minimum Fisher information~\cite{reginatto:1998}, Hall and Reginatto obtain the corresponding nonclassical term from an exact uncertainty principle~\cite{hall:2002}, and the MIW model constructs its interworld potential by discretizing the corresponding term in the quantum average energy~\cite{hall:2014}.

The difference lies in where the functional comes from. In those approaches it is selected by an information-theoretic principle or uncertainty relation, or inherited top-down from the quantum average energy. Here it is what the microscopic interaction~\eqref{eq:vint} becomes when the chain is smoothed, and its coefficient $\hbar^2/(8m)$ is the coupling of that interaction rather than a separate choice.

The same smoothing carries through in three dimensions. Appendix~\ref{app:3d} builds the interaction there and recovers the identical Weizs\"acker functional.

\subsection{Continuity}

We now consider regimes in which the companion-particle number is conserved. No particles are created or destroyed. Introducing a velocity field $v(x,t)$ for the medium, we express this as
\begin{equation}
    \partial_t P + \partial_x (P v) \;=\; 0.
    \label{eq:continuity}
\end{equation}
This continuity equation restricts the smoothed description to nonrelativistic settings. Relativistic regimes with particle creation lie outside its scope.

\subsection{Energy Conservation}

The total energy of the medium consists of the interaction energy~\eqref{eq:vint-smooth} and the kinetic energy of its flow:
\begin{equation}
    E[P, v] \;=\; \int \dd x\, P\!\left[\tfrac{1}{2} m v^2 + V(x)\right] \;+\; \frac{\hbar^2}{8m}\int \dd x\, \frac{(\partial_x P)^2}{P},
\end{equation}
where $V(x)$ is any external potential acting on the particles. To identify the interaction energy contribution, we rewrite the last integral in the form $\int \dd x\, P(x)\, U(x)$. A single integration by parts trades the integrand for itself minus a total derivative,
\begin{equation}
    \int \dd x\,\frac{(\partial_x P)^2}{P}
    \;=\; -\int \dd x\, P\,\partial_x\!\left(\frac{\partial_x P}{P}\right)
    \;=\; \int \dd x\left[\frac{(\partial_x P)^2}{P} - \partial_x^2 P\right],
    \label{eq:integration-by-parts}
\end{equation}
valid whenever $\partial_x P \to 0$ at the boundaries. Note that the microscopic theory needs no such condition: the discrete interaction is well defined whatever happens at the ends of the chain, whereas the \SE{} derived below holds only for the $\partial_x P \to 0$ class of solutions.

Factoring out $P$ puts the interaction energy in the desired form
\begin{equation}
    \frac{\hbar^2}{8m}\int \dd x\,\frac{(\partial_x P)^2}{P}
    \;=\; \int \dd x\, P(x)\, U(x),
    \qquad
    U(x) \;=\; -\frac{\hbar^2}{4m}\!\left[\frac{\partial_x^2 P}{P} - \frac{1}{2}\frac{(\partial_x P)^2}{P^2}\right].
    \label{eq:medium-potential}
\end{equation}
Demanding $\dd E/\dd t = 0$ for arbitrary perturbations, together with continuity, then fixes the evolution of $v$:
\begin{equation}
    m\,\partial_t v + \partial_x\!\left[\tfrac{1}{2} m v^2 + V(x) + U(x)\right] \;=\; 0.
    \label{eq:energy-conservation}
\end{equation}
This is of Hamilton--Jacobi type: the flow velocity is driven by gradients of an effective potential that includes the medium's own contribution $U$.

\subsection{The Schr\"odinger Equation}

The continuity equation~\eqref{eq:continuity} and the energy-conservation equation~\eqref{eq:energy-conservation} form a closed system for $P$ and $v$. Writing $v = \partial_x S / m$ for a scalar field $S(x,t)$, and combining $P$ and $S$ into the single complex field
\begin{equation}
    \psi(x,t) \;\equiv\; \sqrt{P(x,t)}\; \ee^{\ii S(x,t)/\hbar},
\end{equation}
the two real equations become equivalent to the single complex equation
\begin{equation}
    \ii\hbar\, \partial_t \psi \;=\; -\frac{\hbar^2}{2m}\,\partial_x^2 \psi + V(x)\,\psi.
    \label{eq:se}
\end{equation}
This is the Schr\"odinger equation. It is not postulated here, but emerges as the most compact description of the companion-medium dynamics, with $\psi$ playing the role of a bookkeeping device for $P$ and $v$ rather than a fundamental object.

Concerning the last step, Wallstrom observed that the pair~\eqref{eq:continuity} and~\eqref{eq:energy-conservation} admits more solutions than Eq.~\eqref{eq:se}: only the velocity $v = \partial_x S/m$ appears in them, so $S$ is fixed by the dynamics up to an additive constant on each connected region where $P \neq 0$~\cite{wallstrom:1994, reddiger:2023}. Recovering the \SE{} from this equation pair therefore requires a quantization condition, and in the Madelung and stochastic formulations that condition is imposed by hand.

The companion chain supplies it instead. A node is a link carrying a domain wall, which occurs only at a polarity flip. Since $s_k \in \{+1,-1\}$, a finite chain has an integer number of walls. This medium property supplies the integer that labels the states of Section~\ref{sec:examples}: the number of flips selects the stationary state found by relaxation.

\section{Conclusions}
\label{sec:conclusions}

We asked whether particles that are localized at definite positions and obey Newton's equations under an ordinary interaction force can have a coarse-grained description that is the \SE{}. The answer offered here rests on the interaction~\eqref{eq:vint} alone. Every companion compares the signed inverse gaps to its two neighbors, and one coupling constant sets what their mismatch costs.

Both routes from it lead to the same place. Integrated directly, chains of a few dozen to a few hundred particles relax into the stationary densities of the box and the harmonic oscillator and develop double-slit fringes. Smoothed into a density and a velocity field, and constrained by nothing beyond particle-number conservation and energy conservation, the same interaction yields the \SE{}~\eqref{eq:se}.

Several features of this derivation are worth noting:

\begin{itemize}
    \item \textbf{No guidance equation was imposed.} The medium's motion follows from Newton's equations alone.
    \item \textbf{The \SE{} inherits a fixed particle number.} The continuity equation~\eqref{eq:continuity} follows from smoothing a medium of fixed companion count. Particle creation remains possible for the chain.
    \item \textbf{The \SE{} inherits a boundary condition.} Passing from the interaction energy to the per-particle potential $U$ requires $\partial_x P \to 0$ at the ends of the medium. The chain itself needs no such condition.
    \item \textbf{$\psi$ is derived, not fundamental.} It is a compact rewriting of two real fields.
\end{itemize}

The two results announced in the introduction stand as follows:

\begin{itemize}
    \item \textbf{A binary polarity supplies the quantization condition.} A finite chain of binary labels carries an integer number of domain walls. The relaxed chains of Figures~\ref{fig:box} and \ref{fig:hosc} reach the excited eigenstates on that basis alone, whereas the ensemble constructions~\cite{hall:2014, sebens:2015, herrmann:2017} must impose the condition~\cite{wallstrom:1994}.
    \item \textbf{Energy transitions have a mechanism.} With the domain-end energy counted, the relaxed box states match the Schr\"odinger spectrum (Section~\ref{sec:walls}). Moving between them means moving a wall, which is cheapest at a boundary and migrates inward one flip at a time. The medium therefore favors slow, resonant driving over impulsive input of the same energy.
\end{itemize}

\appendix

\section{The Interaction in Three Dimensions}
\label{app:3d}

The construction of Section~\ref{sec:interaction} used the ordering of the companion particles along a line: each particle $k$ compares the inverse distances $1/(x_{k+1}-x_k)$ and $1/(x_k-x_{k-1})$ to its two neighbors. In three dimensions no such ordering exists, but the notion of a neighbor survives. Each companion particle at position $\mathbf{x}_k \in \mathbb{R}^3$ occupies a cell, the region of space closer to it than to any other particle, and its neighbors $N(k)$ are the particles whose cells share a face with its own, $\mathcal{N}_k$ in number. The cells serve only to single out the neighbors. The interaction itself is built, as in one dimension, from inverse distances
\begin{equation}
    u_{kj} \;\equiv\; \frac{s_k\, s_j}{|\mathbf{x}_j - \mathbf{x}_k|},
    \qquad j \in N(k).
    \label{eq:inv-dist}
\end{equation}
This is the three-dimensional analogue of the signed inverse gaps $s_k s_{k+1}/(x_{k+1}-x_k)$.

The same requirements as in one dimension apply: the interaction must vanish when the medium is homogeneous, and must diverge when one neighbor approaches relative to the others. As before, squared differences of inverse distances satisfy both. There, each particle compares its two links. Here, it compares the links to its neighbors pairwise:
\begin{equation}
    \boxed{\;
        V_\mathrm{int} \;=\; g \sum_k
        \sum_{\substack{j < j' \\ j,\,j' \in N(k)}} \bigl( u_{kj} - u_{kj'} \bigr)^2,
        \;}
    \label{eq:vint-3d}
\end{equation}
where the inner sum runs over the unordered pairs of neighbors of $k$ and $g$ is the coupling constant. As in one dimension, $g$ is constrained only to carry dimensions of energy $\times$ length-squared.

In one dimension the cell of particle $k$ is bounded by the midpoints to its two neighbors, so $N(k) = \{k-1, k+1\}$ and the shell contains the single pair $u_{k,k+1} = s_k s_{k+1}/(x_{k+1}-x_k)$, $u_{k,k-1} = s_{k-1} s_k/(x_k-x_{k-1})$. That is, Eq.~\eqref{eq:vint-3d} reduces \emph{exactly} to the interaction~\eqref{eq:vint} of the main text upon the identification $g = \hbar^2/(8m)$.

We now take the smooth limit in a \emph{locally regular phase}: every particle has the same number $\mathcal{N}$ of neighbors, the shells are approximately isotropic, and the polarity is uniform over the region being smoothed. There the distance between neighbors is set by the local density, $|\mathbf{x}_j - \mathbf{x}_k| = \beta P^{-1/3}$ with $\beta$ an order-one constant of the packing, so that $u_{kj} = P^{1/3}/\beta$, evaluated at the midpoint of the link. A pair difference then measures the variation of density across the neighbor shell,
\begin{equation}
    u_{kj} - u_{kj'} \;\approx\; \frac{1}{\beta}\, \nabla P^{1/3} \cdot \frac{\mathbf{x}_j - \mathbf{x}_{j'}}{2}.
\end{equation}
Averaging its square over an isotropic shell of radius $\beta P^{-1/3}$ gives, per pair,
\begin{equation}
    \bigl\langle (u_{kj}-u_{kj'})^2 \bigr\rangle \;=\; \frac{1}{6}\, P^{-2/3}\, |\nabla P^{1/3}|^2,
    \label{eq:pair-average}
\end{equation}
where the packing constant $\beta$ cancels between the inverse link lengths and the shell radius.

Inserting Eq.~\eqref{eq:pair-average}, $|\nabla P^{1/3}|^2 = \tfrac{1}{9}\, P^{-4/3}\, |\nabla P|^2$, $\mathcal{N}(\mathcal{N}-1)/2$ pairs per particle, and $\sum_k \to \int \dd^3x\, P$ for one particle per volume $1/P$, into Eq.~\eqref{eq:vint-3d} results in
\begin{equation}
    V_\mathrm{int} \;\longrightarrow\;
    f\, g \int \dd^3x\, \frac{|\nabla P|^2}{P},
    \qquad
    f \;=\; \frac{\mathcal{N}(\mathcal{N}-1)}{108}.
    \label{eq:vint-3d-smooth}
\end{equation}
Here $f$ is a number set by the coordination of the shell, e.g., $f = 11/9$ for the $\mathcal{N} = 12$ of close packing. The measured value of the reduced Planck constant $\hbar$ is \emph{defined} by this coefficient:
\begin{equation}
    f\, g \;\equiv\; \frac{\hbar^2}{8m}.
    \label{eq:hbar-def}
\end{equation}
Neither $f$ nor $g$ is separately observable, since experiment reaches only their product.

The integral in Eq.~\eqref{eq:vint-3d-smooth} is the three-dimensional form of the von Weizs\"acker density-gradient correction to the local Fermi-gas kinetic energy, introduced in nuclear-mass theory~\cite{weizsacker:1935}. From here the derivation of Section~\ref{sec:schrodinger} goes through with $\partial_x \to \nabla$. The force of the medium is $-\nabla U$ with
\begin{equation}
    U(\mathbf{x}) \;=\; -\frac{\hbar^2}{4m}\!\left[\frac{\nabla^2 P}{P} - \frac{1}{2}\frac{|\nabla P|^2}{P^2}\right],
\end{equation}
and with $\mathbf{v} = \nabla S / m$ and $\psi = \sqrt{P}\,\ee^{\ii S/\hbar}$ the pair of real equations combines into the \SE{}
\begin{equation}
    \ii\hbar\, \partial_t \psi \;=\; -\frac{\hbar^2}{2m}\,\nabla^2 \psi + V(\mathbf{x})\,\psi
\end{equation}
in three dimensions.

\section*{Code Availability}

The simulations of Sections~\ref{sec:examples} and \ref{sec:walls} are reproduced by the code at \url{https://github.com/demaheim/quantum-from-newton}, which also provides the interaction~\eqref{eq:vint} and its gradient as a small library for further use.

\bibliographystyle{apsrev4-2}
\begingroup
\small
\emergencystretch=3em
\bibliography{micro-force}
\endgroup

\end{document}